\documentclass[preprint]{JHEP3} 
\usepackage{graphicx}

\usepackage{amsmath}

\newcommand{\cA}{{\cal A}}
\newcommand{\cD}{{\cal D}}
\newcommand{\cF}{{\cal F}}

\newcommand{\cN}{{\cal N}}
\newcommand{\cQ}{{\cal Q}}
\newcommand{\cU}{{\cal U}}

\newcommand{\hatbnu}{\widehat{\boldsymbol {\nu}}}
\newcommand{\vn}{ {\bf n} }
\newcommand{\uvn}{\underline{\vn}}
\newcommand{\uhatbnu}{\underline{\hatbnu}}
\newcommand{\cAb}{{\overline{\cal A}}}
\newcommand{\cFb}{{\overline{\cal F}}}
\newcommand{\cDb}{{\overline{\cal D}}}
\newcommand{\cUb}{{\overline{\cal U}}} 
\newcommand{\Boxb}{{\overline{\Box}}} 
\newcommand{\Tr}{\;{\rm Tr}}
\newcommand{\hf}{\frac{1}{2}}
\newcommand{\qtr}{\frac{1}{4}}

\def\phib{{\overline{\phi}}}
\def\varphib{{\overline{\varphi}}}
\def\etab{{\overline{\eta}}}
\def\psib{{\overline{\psi}}}
\def\kappab{{\overline{\kappa}}}
\def\thetab{{\overline{\theta}}}
\def\nn{\nonumber}

\def\bec{\begin{center}}
\def\eec{\end{center}}
\def\beq{\begin{equation}}
\def\eeq{\end{equation}}
\def\bea{\begin{eqnarray}}
\def\eea{\end{eqnarray}}

\title{Lattice formulation of three-dimensional ${\cal N}=4$ gauge theory with fundamental matter fields}
\preprint{LA-UR-13-24744}

\author{Anosh Joseph \\
Theoretical Division, Los Alamos National Laboratory, Los Alamos, NM 87545, USA}

\abstract{We construct lattice action for three-dimensional ${\cal N} = 4$ supersymmetric gauge theory with matter fields in the fundamental representation.}


\begin{document}

\section{Introduction}
\label{sec:intro}

There has been a lot of progress in recent years to study several classes of supersymmetric gauge theories on the lattice (see \cite{Kaplan:2003uh, Giedt:2006pd, Catterall:2009it, Joseph:2011xy} for a set of reviews) including the well known $\cN=4$ super Yang-Mills (SYM) that takes part in the AdS/CFT correspondence. Supersymmetric gauge theories are fascinating from a variety of view points; as toy models for studying theories such as QCD, as candidate theories of BSM physics and as theories providing insight into quantum gravity through the gauge/gravity correspondence. These theories exhibit many interesting features at the non-perturbative level, for example, dynamical supersymmetry breaking, and they serve as motivation to study lattice versions of such theories. There have been two distinct formulations to construct certain supersymmetric gauge theories on the lattice while preserving a subset of the continuum supersymmetries \cite{Kaplan:2002wv, Cohen:2003xe, Cohen:2003qw, Kaplan:2005ta, Catterall:2004np, Catterall:2005fd, Catterall:2006jw, Catterall:2005eh}. These lattice theories possess exact supersymmetries, on the contrary to other approaches where supersymmetry only emerges in the continuum limit \cite{Feo:2002yi, Elliott:2005bd}. See \cite{Sugino:2003yb, Sugino:2004qd, D'Adda:2005zk, D'Adda:2007ax, Kanamori:2008bk, Hanada:2009hq, Hanada:2010kt, Hanada:2010gs, Hanada:2011qx} for other recent complementary approaches to the problem of exact lattice supersymmetry.

One approach in constructing supersymmetric lattices is the method of orbifolding. The lattice theory is constructed by imposing the technique of orbifold projection from a parent matrix theory with appropriate symmetries. Substituting the variables of the theory after orbifold projection into the matrix theory gives rise to the desired lattice action \cite{Kaplan:2002wv, Cohen:2003xe, Cohen:2003qw, Kaplan:2005ta}.

The other approach in formulating supersymmetric gauge theories on the lattice involves rewriting of the fermions of the original theory as twisted fermions (anti-symmetric tensor fields) and mapping them to p-cells, p$=0,1,2,3 \cdots$, of the lattice. The lattice action is constructed directly from the continuum twisted theory using the method of geometric discretization. The lattice theory preserves a scalar supercharge at finite lattice spacing \cite{Catterall:2004np, Catterall:2005fd, Catterall:2006jw, Catterall:2005eh}.   

These two formulations appear different from the starting point but the lattices they give are identical \cite{Unsal:2006qp, Catterall:2007kn, Catterall:2009it}. The reason for this is that in the twisting approach the fields are rewritten as representations of the twisted symmetry group, which is the diagonal subgroup of the product of the Euclidean rotation and R-symmetry groups. In the orbifold approach, the placement of orbifold projected variables on the lattice is determined by their charges under the same diagonal subgroup.

Supersymmetric lattice gauge theories have been constructed mostly for pure supersymmetric Yang-Mills theories. There have been a few extensions of these formulations by incorporating matter fields - in Ref. \cite{Endres:2006ic} Endres and Kaplan have constructed a lattice theory for two-dimensional $\cN=(2, 2)$ supersymmetric gauge theory with matter fields in the adjoint representation; in Refs. \cite{Giedt:2006dd, Giedt:2011zza} Giedt has generalized the orbifold lattice formulation of Ref. \cite{Endres:2006ic} to two-dimensional super-QCD with eight supercharges and matter in bi-fundamental representation. The extension, which is relevant to this work, is the two-dimensional lattice constructed by Matsuura in Ref. \cite{Matsuura:2008cfa} for $\cN=(2, 2)$ supersymmetric gauge theory with matter fields in the fundamental representation. In this paper, we write down lattice action for three-dimensional $\cN=4$ supersymmetric lattice gauge theory with fundamental matter. As a side result we also formulate a lattice quiver gauge theory with three-dimensional $\cN=4$ supersymmetry. The quiver theory contains matter fields in the bi-fundamental representation of gauge group $U(N_1) \times U(N_2)$.

The organization of this paper is as follows. In Sec. \ref{sec:3dsym}, we introduce the method of twisting and write down the action of three-dimensional $\cN=4$ SYM using twisted fields. In Sec. \ref{sec:3d-SYM-matter} we construct three-dimensional $\cN=4$ gauge theory with matter fields transforming in the adjoint representation of the gauge group and then rewrite this theory such that matter fields in the fundamental representation are included. We construct the lattice version of three-dimensional twisted $\cN=4$ SYM using the method of geometric discretization in Sec. \ref{sec:latt-form-3dSYM}. The formulation of lattice action for three-dimensional $\cN=4$ SYM with fundamental matter is discussed in Sec. \ref{sec:latt-form-3dSYM-matter}. We discuss the formulation of $\cN=4$ quiver lattice gauge theory containing matter in the bi-fundamental representation of gauge group $U(N_1) \times U(N_2)$ in appendix \ref{sec:latt-quiver}. In Sec. \ref{sec:renorm-simulation} we discuss the renormalization and simulation of three-dimensional $\cN=4$ lattice gauge theory with fundamental matter fields. We end with conclusions and discussion in Sec. \ref{sec:conclusion}.

\section{Three-dimensional $\cN=4$ SYM}
\label{sec:3dsym}

Since we are interested in formulating three-dimensional $\cN=4$ gauge theories on a Euclidean spacetime lattice we will consider only Euclidean versions of gauge theories in our discussion. The three-dimensional Euclidean $\cN=4$ SYM, the starting point of our constructions, can be obtained by dimensionally reducing six-dimensional Euclidean $\cN=1$ SYM. The six-dimensional theory has a gauge field and two independent Weyl spinors. All fields of the theory are in the adjoint representation of gauge group, which we take as $U(N)$ in this paper. After reducing to three dimensions the Weyl spinors split into two independent four-component complex spinors and the gauge field reduces to a three-dimensional gauge field and three real scalars. The global symmetry group of the three-dimensional theory is $SU(2)_E \times SU(2)_R \times SU(2)_N$, where $SU(2)_E$ is the Euclidean rotation group in three dimensions, $SU(2)_R$ is the R-symmetry group of the six-dimensional theory and $SU(2)_N$ is the internal Euclidean rotation group arising from the decomposition $SO(6) \rightarrow SU(2)_E \times SU(2)_N$. Since our goal in this paper is to construct three-dimensional $\cN=4$ lattice gauge theory with matter fields, we will rewrite the fields and supersymmetries in a way more suitable for lattice discretization. As mentioned in the previous section, there are two approaches immediately available to us - the method of orbifolding \cite{Kaplan:2002wv, Kaplan:2003uh, Cohen:2003xe, Cohen:2003qw, Kaplan:2005ta} and the technique of topological twisting \cite{Catterall:2004np, Catterall:2005fd, Catterall:2006jw, Catterall:2005eh}. They both ultimately give similar lattices for the theory. In Ref. \cite{Matsuura:2008cfa} Matsuura constructed a two-dimensional $\cN = (2, 2)$ lattice gauge theory with matter fields in the fundamental representation of the gauge group $U(N)$. There, it has been shown that identical lattice theories can be obtained from the two approaches. In this paper, we focus on the topological twisting approach and use topologically twisted version of the three-dimensional $\cN=4$ SYM theory to write down the action of the lattice gauge theory. We briefly describe the twisting process of three-dimensional $\cN=4$ SYM below.

\subsection{Topological twisting: A lattice compatible relabeling}
\label{sec:twisitng}

The key idea of twisting is to decompose the fields and supersymmetries of the three-dimensional $\cN=4$ SYM in terms of representations of a newly defined rotation group instead of the original Euclidean rotational symmetry $SU(2)_E$. The new rotation group, $SU(2)^\prime$, is called the twisted rotation group and it is defined as the diagonal subgroup of $SU(2)_E$ and $SU(2)_N$,
\beq
SU(2)'={\rm diag}\Big(SU(2)_E \times SU(2)_N\Big).
\eeq
This particular twist of the theory is known as the Blau-Thompson twist \cite{Blau:1996bx}. After twisting, the field content of the original theory becomes representations of the twisted rotation group. It should be noted that in flat Euclidean spacetime the process of twisting is nothing but a change of variables of the original theory. 

The twisting process gives rise to the following spectrum of the twisted theory: a three-dimensional gauge field $A_m$, $m=1,2,3$; a vector $B_m$ composed of three scalars of the untwisted theory; and eight p-form fermions, p $= 0, 1, 2, 3$, which we conveniently represent as $\{\eta, \psi_m, \chi_{mn}, \theta_{mnr}\}$. 

The supercharges of the theory undergo a decomposition similar to that of the fermions after twisting. They take the form $\{\cQ, \cQ_m, \cQ_{mn}, \cQ_{mnr}\}$ and are called the twisted supercharges. We can write down the original supersymmetry algebra in terms of the twisted supercharges
\beq
\cQ^2 = 0,\; \{\cQ,\cQ_m\} = p_m,\; \cdots
\eeq
The first equation shows that the process of twisting produces a nilpotent scalar supercharge $\cQ$. Since the scalar supercharge does not produce any infinitesimal translations, we can transport this subalgebra to the lattice. The second equation of the twisted algebra has the following interpretation: The momentum is the $\cQ$-variation of something, which makes plausible the statement that the energy-momentum tensor, and hence the entire action can be written in a $\cQ$-exact form. This implies that we can construct a lattice action in a $\cQ$-exact form and it is trivially invariant under the scalar supercharge. Thus the process of twisting can be used to construct lattice action that respects at least one supersymmetry exactly on the lattice. We also note that the lattice theories constructed using twisted fermions are free from the fermion doubling problem, since they are geometric in nature (p-forms) and thus can be mapped one-to-one on to the lattice from continuum \cite{Rabin:1981qj, Becher:1982ud, Banks:1982iq, Aratyn:1984bd}.

\subsection{Action and supersymmetries}
\label{sec:action-susy}

The twisted action of the three-dimensional $\cN=4$ SYM takes the following $\cQ$-exact form in the continuum
\bea
S &=&\frac{1}{g^2}\cQ \int d^3x \Tr \left(\chi_{mn}[\cD_m, \cD_n] + \eta \left[\cDb_m,\cD_m\right] + \hf\eta d + B_{mnr}\cDb_r \chi_{mn}\right),
\label{action1}
\eea
with $g$ the coupling constant of the theory. Since the twisted theory contains two vector fields, $A_m$ and $B_m$, it is natural to combine them to form a complex gauge field $\cA_m = A_m + i B_m$. Thus the degrees of freedom of the twisted theory are just the twisted fermions $\{\eta, \psi_m, \chi_{mn}, \theta_{mnr}\}$ previously described and the complex gauge field $\cA_m$.

The theory contains complexified covariant derivatives and they are defined by
\bea
\label{eq:cov-der-adj1}
\cD_m ~\cdot &=& \partial_m \cdot + ~[\cA_m, ~\cdot~ ] = \partial_m \cdot + ~[A_m + iB_m, ~\cdot~ ],\\
\label{eq:cov-der-adj2}
\cDb_m ~\cdot &=& \partial_m \cdot + ~[\cAb_m, ~\cdot~ ] = \partial_m \cdot + ~[A_m - i B_m, ~\cdot~ ].
\eea
The complexification of gauge field also results in complexified field strength $\cF_{mn} = [\cD_m, \cD_n]$ and $\cFb_{mn} = [\cDb_m, \cDb_n]$. All fields take values in the adjoint representation of the gauge group $U(N)$. Although the theory contains a complexified gauge field and field strength, it possesses only the usual $U(N)$ gauge-invariance corresponding to the real part of the gauge field. Fields $d$ and $B_{mnr}$ are auxiliaries introduced to render the scalar supersymmetry $\cQ$ nilpotent off-shell.

The scalar supersymmetry acts on the fields the following way
\begin{align}
\label{eq:susy}
\cQ \cA_m& =\psi_m,&
\cQ \cAb_m& =0,\\
\cQ \psi_m& =0,&
\cQ \chi_{mn}& =-\cFb_{mn},\\
\cQ \eta& =d,&
\cQ d& =0,\\
\cQ B_{mnr}& =\theta_{mnr},&
\cQ \theta_{mnr}& =0.
\end{align}

After performing the $\cQ$-variation, integrating out the auxiliary fields and using the Bianchi identity, $\epsilon_{mnr}\cDb_r \cFb_{mn} = 0$, the action becomes
\bea
\label{action}
S &=& \frac{1}{g^2}\int d^3 x~\Tr \Big(-\cFb_{mn}\cF_{mn} + \hf[\cDb_m, \cD_m]^2 -\chi_{mn}(\cD_m\psi_n - \cD_n\psi_m) \nn \\
&&- \psi_m\cDb_m\eta - \theta_{mnr}\cDb_{\left[r \right.} \chi_{\left.mn\right]}\Big).
\eea
The terms appearing in the bosonic piece of the action can be written in the following form exposing the $B_m$ dependence explicitly
\bea
\cFb_{mn}\cF_{mn} &=& (F_{mn} - [B_m, B_n])(F_{mn} - [B_m, B_n]) + (D_{\left[m\right.}B_{\left.n\right]})
(D_{\left[m\right.}B_{\left.n\right]}),\nn \\
\hf\left[\cDb_m, \cD_m\right]^2 &=& -2\left(D_m B_m\right)^2,
\eea
where $F_{mn}$ and $D_m$ denote the usual field strength and covariant derivative depending on the real part of the gauge field $\cA_m$. 

\section{Three-dimensional $\cN=4$ SYM with matter fields}
\label{sec:3d-SYM-matter}

In this section we construct extensions of the three-dimensional $\cN=4$ SYM described in Sec. \ref{sec:3dsym} with the inclusion of matter fields in the adjoint and fundamental representations of the gauge group. We construct such theories by dimensionally reducing the twisted version of four-dimensional $\cN=4$ SYM.

The four-dimensional $\cN=4$ SYM contains a gauge field, six real scalars and four Weyl fermions (gauginos) in the adjoint representation of the gauge group. The twist appropriate for lattice construction of this theory is known as Marcus twist \cite{Marcus:1995mq}. (See Ref. \cite{Catterall:2005fd} for details of the lattice construction of four-dimensional twisted $\cN=4$ SYM.) After twisting, the fermionic degrees of freedom are encoded in the twisted p-form fields $(\eta, \psi_\mu, \chi_{\mu \nu}, \theta_{\mu \nu \rho}, \kappa_{\mu \nu \rho \sigma})$, with Greek indices running from $1, \cdots, 4$, and the bosonic degrees of freedom are encoded in a complexified gauge field $\cA_\mu$ and two scalars $\phi$ and $\phib$.

The action of the four-dimensional twisted $\cN=4$ SYM is given by
\bea
\label{eq:4d-twist-action}
S &=& \frac{1}{g_4^2} \int d^4x \Tr \Big(-[\cDb_\mu, \cDb_\nu][\cD_\mu, \cD_\nu] - 2(\cDb_\mu\phib)(\cD_\mu\phi) + \hf \Big([\cDb_\mu, \cD_\mu] + [\phib, \phi]\Big)^2 \nn \\
&&- \psi_\mu \cDb_\mu \eta - \chi_{\mu\nu} (\cD_\mu \psi_\nu - \cD_\nu \psi_\mu) - \frac{1}{3!}\epsilon_{\nu\sigma\lambda\rho} \epsilon_{\nu\alpha\beta\delta} \chi_{\lambda\rho} \cDb_\sigma \theta_{\alpha\beta\delta} \nn \\
&&+ 2\frac{1}{3!}\frac{1}{4!}\epsilon_{\mu \alpha \beta \delta}\epsilon_{\sigma \nu \lambda \rho} \theta_{\alpha \beta \delta} \cD_\mu \kappa_{\sigma \nu \lambda \rho} - \qtr \epsilon_{\mu\nu\lambda\rho} \chi_{\mu\nu} [\phib, \chi_{\lambda\rho}] \nn \\
&&- 2\frac{1}{3!}\epsilon_{\mu\nu\lambda\rho}\theta_{\nu\lambda\rho}[\phi,\psi_\mu] - \frac{1}{4!}\epsilon_{\alpha\beta\delta\sigma} \kappa_{\alpha\beta\delta\sigma} [\phib, \eta] \Big),
\eea
with $g_4$ denoting the four-dimensional coupling constant. 

\subsection{Action with adjoint matter}
\label{sec:action-adj-matter}

We obtain the following form of the action for the three-dimensional theory after dimensionally reducing Eq. (\ref{eq:4d-twist-action}), the action of the four-dimensional twisted $\cN=4$ SYM,
\beq
\label{eq:3d-adj-matter}
S = S_{\bf adj}^{\rm SYM} + S_{\bf adj}^{\rm matter},
\eeq
where
\bea
\label{eq:adj-3d-sym}
S_{\bf adj}^{\rm SYM} &=& \frac{1}{g^2} \int d^3x \Tr~ (-\cFb_{mn}\cF_{mn} + \hf [\cDb_m, \cD_m]^2 - \chi_{mn} (\cD_m \psi_n - \cD_n \psi_m) \nn \\
&&- \psi_m \cDb_m \eta - \theta_{mnr} \cDb_r \chi_{mn} \Big),
\eea
and
\bea
S_{\bf adj}^{\rm matter} &=& \frac{1}{g^2}\int d^3x \Tr~ \Big([\cDb_m, \cD_m]([\varphib, \varphi]+ [\phib, \phi]) -2 (\cDb_m\varphib)(\cD_m\varphi) - 2(\cDb_m\phib)(\cD_m\phi) \nn \\
&&+ 2\psib_m \cD_m \etab - 2 \kappab_{np}\cDb_p\psib_n + 2 \thetab_{npm}\cD_m\kappab_{np} - \eta [\phib, \etab] - \epsilon_{npm} \eta [\varphib, \thetab_{npm}] \nn \\
&&+ \frac{1}{3} \epsilon_{mnr} \theta_{mnr} [\varphi, \etab] - 2\psi_m[\phi, \psib_m] - \frac{1}{3}\epsilon_{npm} \psi_m[\varphi, \kappab_{np}] + \epsilon_{npq} \chi_{pq}[\varphib, \psib_n] \nn \\
&&- \chi_{pq} [\phib, \kappab_{pq}] - 2\theta_{mnr} [\phi, \thetab_{mnr}] + \hf \Big([\varphib, \varphi]+ [\phib, \phi]\Big)^2 - 2[\varphib, \phib][\varphi, \phi]\Big).
\eea

The first piece of the action is the twisted action of the three-dimensional $\cN=4$ SYM described in Sec. \ref{sec:3dsym}. The second piece includes matter fields $\{\phi$, $\phib$, $\varphi$, $\varphib$, $\etab$, $\psib_m$, $\kappab_{mn}$, $\thetab_{mnr}\}$ in the adjoint representation. 

The fields of the three-dimensional theory respect the following scalar supersymmetry transformations
\begin{align}
\label{eq:susy-3d-1}
\cQ \cA_m& =\psi_m,&
\cQ \cAb_m& =0,\\
\cQ \eta& =[\cDb_m, \cD_m] + [\varphib, \varphi] + [\phib, \phi],&
\cQ \psi_m& =0,\\
\cQ \chi_{mn} & =-[\cDb_m, \cDb_n],&
\cQ \theta_{mnr}& = \epsilon_{mnr}[\varphib, \phib],\\
\cQ \phi& =\etab,&
\cQ \phib& =0,\\
\cQ \varphi& =\epsilon_{mnr}\thetab_{mnr},&
\cQ \varphib& =0,\\
\cQ \etab& =0,&
\cQ \psib_m& = \cDb_m\phib,\\
\label{eq:susy-3d-2}
\cQ \kappab_{mn} & = \epsilon_{mnr} \cDb_r\varphib,&
\cQ \thetab_{mnr}& =0.
\end{align}

\subsection{Action with fundamental matter}
\label{sec:action-fnd-matter}

\subsubsection{Side step: A quiver theory with bi-fundamental matter}
\label{subsubsec:action-bi-fnd-matter}

\begin{figure}
\begin{center}
\includegraphics[width=0.4\textwidth]{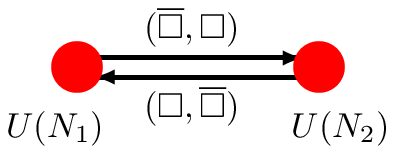}
\end{center}
\caption{\label{fig:quiver_diagram}A quiver diagram for $U(N_1) \times U(N_2)$ gauge theory. The links with arrows represent bi-fundamental fields.}
\end{figure}

We can rewrite the action, Eq. (\ref{eq:3d-adj-matter}), such that the theory becomes a three-dimensional quiver gauge theory with $\cN=4$ supersymmetry. In the case we are interested in, there are two interacting $U(N)$ gauge theories. The SYM multiplets of this quiver gauge theory transform in the adjoint representation of the gauge group $U(N_1) \times U(N_2)$. The two theories interact via matter multiplets in the bi-fundamental representation of $U(N_1) \times U(N_2)$. (See Fig. \ref{fig:quiver_diagram}.) The action of the quiver theory can be decomposed in the following way
\beq
S = S^{\rm SYM}_{({\bf adj}, {\bf 1})} + S^{\rm SYM}_{({\bf 1}, {\bf adj})} + S^{\rm matter}_{(\Box, \Boxb)} + S^{\rm matter}_{(\Boxb, \Box)},
\eeq
with the field content of the theory $\{\cA_m$, $\cAb_m$, $\eta$, $\psi_m$, $\chi_{mn}$, $\theta_{mnr}\}$, $\{\widehat{\cA}_m$, $\widehat{\cAb}_m$, $\widehat{\eta}$, $\widehat{\psi}_m$, $\widehat{\chi}_{mn}$, $\widehat{\theta}_{mnr}\}$, $\{\phi$, $\widehat{\phib}$, $\widehat{\varphi}$, $\varphib$, $\etab$, $\widehat{\psib}_m$, $\kappab_{mn}$, $\widehat{\thetab}_{mnr}\}$ and $\{\widehat{\phi}$, $\phib$, $\varphi$, $\widehat{\varphib}$, $\widehat{\etab}$, $\psib_m$, $\widehat{\kappab}_{mn}$, $\thetab_{mnr}\}$ transforming respectively as $({\bf adj}, {\bf 1})$, $({\bf 1}, {\bf adj})$, $(\Box, \Boxb)$ and $(\Boxb, \Box)$ under $U(N_1) \times U(N_2)$.

We have
\bea
S^{\rm SYM}_{({\bf adj}, {\bf 1})} &=& \frac{1}{g^2} \int d^3x \Tr~\Big(-\cFb_{mn} \cF_{mn} + \hf [\cDb_m, \cD_m]^2 - \chi_{mn} (\cD_m\psi_n - \cD_n\psi_m) \nn \\
&&~~~~~~~~~~~~~~~~~~- \psi_m \cDb_m\eta - \theta_{mnr}\cDb_r \chi_{mn} \Big),
\eea

\bea
S^{\rm SYM}_{({\bf 1}, {\bf adj})} &=& \frac{1}{g^2} \int d^3x \Tr~\Big(-\widehat{\cFb}_{mn} \widehat{\cF}_{mn} + \hf [\widehat{\cDb}_m, \widehat{\cD}_m]^2 - \widehat{\chi}_{mn} (\widehat{\cD}_m\widehat{\psi}_n - \widehat{\cD}_n\widehat{\psi}_m) \nn \\
&&~~~~~~~~~~~~~~~~~~- \widehat{\psi}_m \widehat{\cDb}_m\widehat{\eta} - \widehat{\theta}_{mnr}\widehat{\cDb}_r \widehat{\chi}_{mn} \Big),
\eea

\bea
S^{\rm matter}_{(\Box, \Boxb)} &=& \frac{1}{g^2}\int d^3x \Tr~ \Big([\cDb_m, \cD_m](\varphib \varphi - \widehat{\varphi} \widehat{\varphib} + \widehat{\phib} \widehat{\phi} - \phi \phib) + 2 \widehat{\varphi} \cD_m\cDb_m \widehat{\varphib} \nn \\
&&+ 2 \widehat{\phib} \cDb_m\cD_m \widehat{\phi} + 2\widehat{\psib}_m \cD_m \widehat{\etab} + 2 \widehat{\thetab}_{npm}\cD_m\widehat{\kappab}_{np} \nn \\
&&- 2 \kappab_{np}\cDb_p\psib_n - \eta (\widehat{\phib}\widehat{\etab} - \etab\phib) - \epsilon_{npm} \eta (\varphib \thetab_{npm} - \widehat{\thetab}_{npm}\widehat{\varphib}) \nn \\
&&+ \frac{1}{3} \epsilon_{mnr} \theta_{mnr} (\widehat{\varphi}\widehat{\etab} - \etab\varphi) - 2\psi_m (\phi\psib_m - \widehat{\psib}_m\widehat{\phi}) - \frac{1}{3}\epsilon_{npm} \psi_m (\widehat{\varphi}\widehat{\kappab}_{np} - \kappab_{np}\varphi) \nn \\
&&+ \epsilon_{npq} \chi_{pq} (\varphib\psib_n - \widehat{\psib}_n\widehat{\varphib}) - \chi_{pq} (\widehat{\phib}\widehat{\kappab}_{pq} - \kappab_{pq}\phib) - 2\theta_{mnr}(\phi\thetab_{mnr} - \widehat{\thetab}_{mnr}\widehat{\phi}) \nn \\
&&+ \hf \Big((\varphib\varphi - \widehat{\varphi}\widehat{\varphib})+ (\widehat{\phib}\widehat{\phi} - \phi\phib)\Big)^2 - 2(\varphib\phib - \widehat{\phib}\widehat{\varphib})(\widehat{\varphi}\widehat{\phi} - \phi\varphi)\Big),
\eea

and

\bea
S^{\rm matter}_{(\Boxb, \Box)} &=& \frac{1}{g^2}\int d^3x \Tr~ \Big([\cDb_m, \cD_m](\widehat{\varphib} \widehat{\varphi} - \varphi \varphib + \phib \phi - \widehat{\phi} \widehat{\phib}) + 2 \varphi \cD_m\cDb_m \varphib \nn \\
&&+ 2 \phib \cDb_m\cD_m \phi + 2\psib_m \cD_m \etab + 2 \thetab_{npm}\cD_m\kappab_{np} \nn \\
&&- 2 \widehat{\kappab}_{np}\cDb_p\widehat{\psib}_n - \widehat{\eta} (\phib\etab - \widehat{\etab}\widehat{\phib}) - \epsilon_{npm} \widehat{\eta} (\widehat{\varphib} \widehat{\thetab}_{npm} - \thetab_{npm}\varphib) \nn \\
&&+ \frac{1}{3} \epsilon_{mnr} \widehat{\theta}_{mnr} (\varphi\etab - \widehat{\etab}\widehat{\varphi}) - 2\widehat{\psi}_m (\widehat{\phi}\widehat{\psib}_m - \psib_m\phi) - \frac{1}{3}\epsilon_{npm} \widehat{\psi}_m (\varphi\kappab_{np} - \widehat{\kappab}_{np}\widehat{\varphi}) \nn \\
&&+ \epsilon_{npq} \widehat{\chi}_{pq} (\widehat{\varphib}\widehat{\psib}_n - \psib_n\varphib) - \widehat{\chi}_{pq} (\phib\kappab_{pq} - \widehat{\kappab}_{pq}\widehat{\phib}) - 2\widehat{\theta}_{mnr}(\widehat{\phi}\widehat{\thetab}_{mnr} - \thetab_{mnr}\phi) \nn \\
&&+ \hf \Big((\widehat{\varphib}\widehat{\varphi} - \varphi\varphib)+ (\phib\phi - \widehat{\phi}\widehat{\phib})\Big)^2 - 2(\widehat{\varphib}\widehat{\phib} - \phib\varphib)(\varphi\phi - \widehat{\phi}\widehat{\varphi})\Big).
\eea

There are two types of covariant derivatives appearing in the above expressions, acting respectively on adjoint and bi-fundamental matter. The covariant derivatives for the adjoint matter are given in Eqs. (\ref{eq:cov-der-adj1}) - (\ref{eq:cov-der-adj2}). For a generic bi-fundamental matter field $\phi$ in the representation $(\Box, \Boxb)$ we have the action of the covariant derivative
\beq
\cD_m \phi = \partial_m \phi + \cA_m \phi - \phi \widehat{\cA}_m,
\eeq
with $\cA_m$ and $\widehat{\cA}_m$ the gauge fields for $U(N_1)$ and $U(N_2)$ respectively. The gauge transformation rule for the field $\phi$, under $(G, \widehat{G}) \in U(N_1) \times U(N_2)$, is given by $\phi \rightarrow G \phi \widehat{G}^\dagger$. For a field $\widehat{\phi}$ in the representation $(\Boxb, \Box)$ we have the action of the covariant derivative
\beq
\cD_m \widehat{\phi} = \partial_m \widehat{\phi} + \widehat{\cA}_m \widehat{\phi} - \widehat{\phi} \cA_m,
\eeq
with the rule for gauge transformation: $\widehat{\phi} \rightarrow \widehat{G} \widehat{\phi} G^\dagger$.

The three-dimensional quiver gauge theory with $\cN=4$ supersymmetry constructed here, though not the main result of this paper, is interesting in its own right. This quiver theory construction can be easily transported on to the lattice. The lattice construction of three-dimensional $\cN=4$ quiver $U(N_1) \times U(N_2)$ gauge theory is given in Appendix \ref{sec:latt-quiver}.

\subsubsection{From quiver theory to action with fundamental matter}
\label{subsubsec:action-fnd-matter}

To construct three-dimensional $\cN=4$ lattice SYM theory with matter in the fundamental representation we freeze one of the theories say, the one with the gauge group $U(N_2)$ and also the set of matter fields decorated with hats. After this restriction we have a $U(N_1)$ gauge theory containing matter fields in fundamental representation. Note that the restriction of the fields is not in conflict with supersymmetry. The resultant action is still $\cQ$-invariant.  

The scalar supersymmetry acts on the fields the following way
\begin{align}
\label{eq:susy-3d-1}
\cQ \cA_m& =\psi_m,&
\cQ \cAb_m& =0,\\
\cQ \eta& =[\cDb_m, \cD_m] + \varphib\varphi - \phi\phib,&
\cQ \psi_m& =0,\\
\cQ \chi_{mn} & =-[\cDb_m, \cDb_n],&
\cQ \theta_{mnr}& = \epsilon_{mnr} \varphib\phib,\\
\cQ \phi& =\etab,&
\cQ \phib& =0,\\
\cQ \varphi& =\epsilon_{mnr}\thetab_{mnr},&
\cQ \varphib& =0,\\
\cQ \etab& =0,&
\cQ \psib_m& = \cDb_m\phib,\\
\label{eq:susy-3d-2}
\cQ \kappab_{mn} & = \epsilon_{mnr} \cDb_r\varphib,&
\cQ \thetab_{mnr}& =0.
\end{align}

The action of the three-dimensional $\cN=4$ gauge theory contains two pieces
\beq
S = S^{\rm SYM} + S^{\rm matter},
\eeq
where $S^{\rm SYM}$ contains adjoint fields and the expression is given in Eq. (\ref{eq:adj-3d-sym}). The piece $S^{\rm matter}$ contains matter fields in the fundamental representation. It is given by 

\bea
S^{\rm matter} &=& \frac{1}{g^2}\int d^3x \Tr~ \Big([\cDb_m, \cD_m](\varphib \varphi - \phi \phib) + 2 \varphi \cD_m\cDb_m \varphib + 2 \phib \cDb_m\cD_m \phi \nn \\
&&+ 2\psib_m \cD_m \etab - 2 \kappab_{np}\cDb_p\psib_n + 2 \thetab_{npm}\cD_m\kappab_{np} \nn \\
&&+ \eta \etab \phib - \epsilon_{npm} \eta \varphib \thetab_{npm} - \frac{1}{3} \epsilon_{mnr} \theta_{mnr} \etab \varphi - 2\psi_m \phi \psib_m \nn \\
&&+ \frac{1}{3} \epsilon_{npm} \psi_m \kappab_{np} \varphi + \epsilon_{npq} \chi_{pq} \varphib \psib_n + \chi_{pq} \kappab_{pq} \phib - 2\theta_{mnr} \phi \thetab_{mnr} \nn \\
&&+ \hf \Big(\varphib\varphi - \phi\phib\Big)^2 + \hf \Big(\phib\phi - \varphi\varphib\Big)^2 + 2(\varphib\phib)(\phi\varphi) + 2(\phib\varphib)(\varphi\phi) \Big).
\eea

This theory can be discretized on the lattice. In Sec. \ref{sec:latt-form-3dSYM-matter}, we construct three-dimensional $\cN=4$ lattice SYM theory with matter in the fundamental representation.

\section{Lattice formulation of three-dimensional $\cN=4$ SYM}
\label{sec:latt-form-3dSYM}

\subsection{Geometric discretization}
\label{sec:geom-discrtzn}

\begin{figure}
\begin{center}
\includegraphics[width=0.45\textwidth]{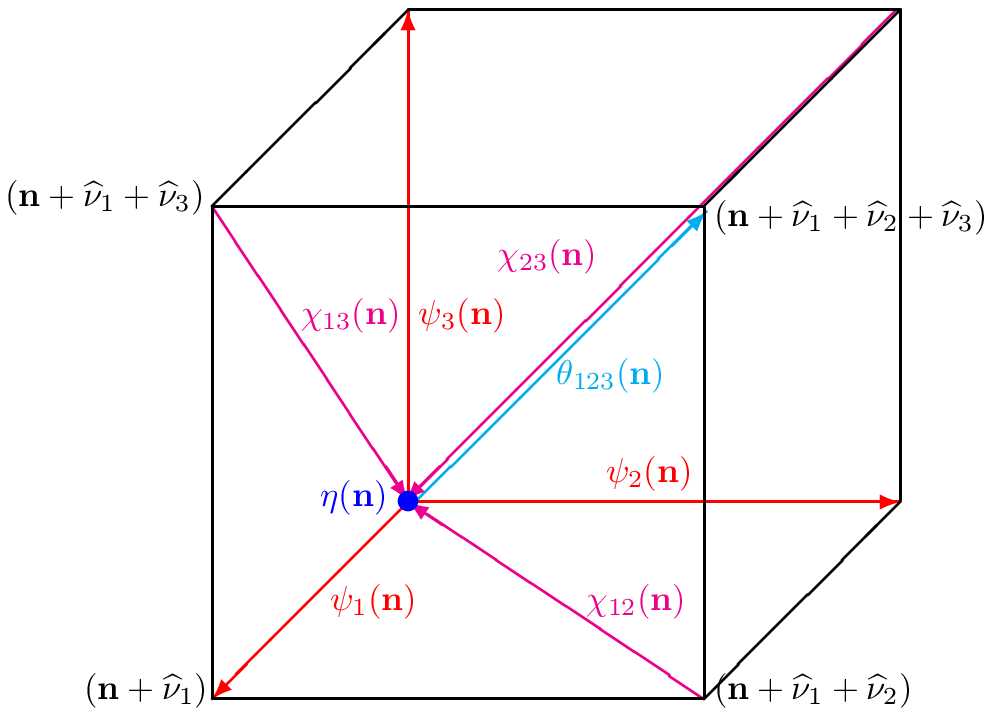}
\end{center}
\caption{\label{fig:3dLattice}The unit cell of three-dimensional $\cN=4$ lattice SYM with orientation assignments for twisted fermionic fields. The bosonic fields $\cU_m$ follow the same orientations and link assignments as that of their superpartners $\psi_m$.}
\end{figure}

The discretization of the twisted theory described in the previous section is straightforward. We use the technique of geometric discretization developed in Refs. \cite{Catterall:2007kn, Damgaard:2007be, Damgaard:2008pa}. We replace the continuum complex gauge field $\cA_m(x), m = 1, 2, 3$, at every point by an appropriate complexified Wilson link $\cU_m(\vn) = e^{\cA_m(\vn)}$. These lattice fields are taken to be associated with unit length vectors in the coordinate directions $\hatbnu_m$ from the site denoted by the integer vector $\vn$ on a three-dimensional hypercubic lattice. Supersymmetric invariance then implies that the fermion fields $\psi_m(\vn)$ lie on the same oriented link as their bosonic superpartners $\cU_m(\vn)$, running from $\vn \to \vn + \hatbnu_m$. The scalar fermion $\eta(\vn)$ is associated with the site $\vn$ of the lattice. The components of the field $\chi_{mn}(\vn), m < n = 1, 2, 3$, are placed on a set of diagonal face links running from $\vn + \hatbnu_m + \hatbnu_n \to \vn$. The 3-form field $\theta_{mnr}(\vn)$ is placed on the body diagonal running from $\vn \to \vn + \hatbnu_m + \hatbnu_n + \hatbnu_r$. The unit cell and the field orientations of the three-dimensional theory are given in Fig. \ref{fig:3dLattice}.

The lattice prescription for topologically twisted theory is the following: lattice variables $\cU_a(\vn)$, $\cUb_a(\vn)$, $\{f^{(+)}_{a_1 \cdots a_p}(\vn) \}$,  $\{f^{(-)}_{a_1 \cdots a_p}(\vn) \}$ live on links $(\vn, \vn+\hatbnu_a)$, $(\vn + \hatbnu_a, \vn)$, $(\vn, \vn+\hatbnu_{a_1}+\cdots+\hatbnu_{a_p})$ and $(\vn+\hatbnu_{a_1}+\cdots+\hatbnu_{a_p}, \vn)$ respectively. A site variable $f(\vn)$ lives on a degenerate link $(\vn, \vn)$. 

We can write down the gauge transformation rules for the (adjoint) lattice fields respecting the p-cell and orientation assignments on the lattice. For $G(\vn) \in U(N)$, we have the following gauge transformation prescription \cite{Aratyn:1984bd, Damgaard:2008pa}
\bea
\label{eq:gauge-lattice-1}
\cU_a(\vn) &\rightarrow& G(\vn) \cU_a(\vn) G^{\dagger}(\vn + \hatbnu_a), \\ 
\cUb_a(\vn) &\rightarrow& G(\vn + \hatbnu_a) \cUb_a(\vn)G^{\dagger}(\vn), \\ 
\{f^{(+)}_{a_1 \cdots a_p}(\vn) \} &\rightarrow& G(\vn) \{f^{(+)}_{a_1 \cdots a_p}(\vn) \} G^{\dagger}(\vn + \hatbnu_{a_1} + \cdots+\hatbnu_{a_p}), \\ 
\label{eq:gauge-lattice-2}
\{f^{(-)}_{a_1 \cdots a_p}(\vn) \} &\rightarrow& G(\vn + \hatbnu_{a_1} + \cdots+\hatbnu_{a_p} )\{f^{(-)}_{a_1 \cdots a_p}(\vn) \} G^{\dagger}(\vn).
\eea

We need to describe how continuum covariant derivatives are to be replaced by covariant difference operators. The covariant derivatives $\cD_a$ ($\cDb_a$) in the continuum become forward and backward covariant differences $\cD^{(+)}_a~(\cDb^{(+)}_a)$ and $\cD^{(-)}_a~(\cDb^{(-)}_a)$, respectively. The forward covariant difference operator acts on the lattice fields $f^{(\pm)}_{a_1 \cdots a_p}(\vn)$ in the following way:
\bea
\label{eq:cov-diff-adj-1}
&&\cD_b^{(+)}f^{(+)}_{a_1 \cdots a_p}(\vn) \equiv \cU_b(\vn)f^{(+)}_{a_1 \cdots a_p}(\vn + \hatbnu_b)-f^{(+)}_{a_1 \cdots a_p}(\vn) \cU_b(\vn+\hatbnu),~~~~~~~~ \\
&&\cD_b^{(+)}f^{(-)}_{a_1 \cdots a_p}(\vn) \equiv \cU_b(\vn+\hatbnu)f^{(-)}_{a_1 \cdots a_p}(\vn + \hatbnu_b)-f^{(-)}_{a_1 \cdots a_p}(\vn) \cU_b(\vn),~~~~~~~~ \\
&&\cDb_b^{(+)}f^{(+)}_{a_1 \cdots a_p}(\vn) \equiv f^{(+)}_{a_1 \cdots a_p}(\vn + \hatbnu_b)\cUb_b(\vn+\hatbnu)-\cUb_b(\vn)f^{(+)}_{a_1 \cdots a_p}(\vn),~~~~~~~~ \\
&&\cDb_b^{(+)}f^{(-)}_{a_1 \cdots a_p}(\vn) \equiv f^{(-)}_{a_1 \cdots a_p}(\vn + \hatbnu_b)\cUb_b(\vn)-\cUb_b(\vn+\hatbnu)f^{(-)}_{a_1 \cdots a_p}(\vn),~~~~~~~~
\eea
where we have defined $\hatbnu = \sum_{i=1}^p \hatbnu_{a_i}$. 

The action of the backward covariant difference operator on the lattice fields is given by: 
\bea
\cD_b^{(-)}f^{(\pm)}_{a_1 \cdots a_p}(\vn) &\equiv& \cD_b^{(+)}f^{(\pm)}_{a_1 \cdots a_p}(\vn-\hatbnu_b), \\
\label{eq:cov-diff-adj-2}
\cDb_b^{(-)}f^{(\pm)}_{a_1 \cdots a_p}(\vn) &\equiv& \cDb_b^{(+)}f^{(\pm)}_{a_1 \cdots a_p}(\vn-\hatbnu_b).
\eea

These expressions are determined by the two requirements that they reduce to the corresponding continuum results for the adjoint covariant derivative in the naive continuum limit and that they transform under gauge transformations like the corresponding lattice link field carrying the same indices. As a result, the terms in the lattice action correspond to gauge-invariant closed loops. 

The lattice field strength is given by the expression $\cF_{mn}(\vn) = \cD^{(+)}_m \cU_n(\vn)$. We see that it is automatically antisymmetric in its indices and also it transforms like a lattice 2-form.

\subsection{Lattice action}
\label{sec:latt-action}

Having the above prescription for geometric discretization in hand, we can write down the supersymmetric and gauge-invariant lattice action of the three-dimensional $\cN=4$ SYM
\bea
\label{eq:3d-latt-action}
S &=& \frac{1}{g^2}\sum_{\vn} \Tr~ \Big(-\cFb_{mn}(\vn)\cF_{mn}(\vn) + \frac{1}{2}\Big(\cDb_m^{(-)}\cU_m(\vn)\Big)^2 \nn \\
&&~~~~~~~~~~~~~~~~~~-\chi_{mn}(\vn)\cD_{[m}^{(+)}\psi_{n]}(\vn) - \eta(\vn) \cDb_m^{(-)}\psi_m(\vn) \nn \\
&&~~~~~~~~~~~~~~~~~~- \theta_{mnr}(\vn)\cDb_{[r}^{(+)}\chi_{mn]}(\vn)\Big).~~~~~~
\eea

The bosonic part of the action is
\bea
\label{eq:3d-bosonic-adj}
S_B &=& \frac{1}{g^2}\sum_{\vn} \Tr~ \Big[-\Big(\overline{\cD_m^{(+)}\cU_n(\vn)}\Big)\Big(\cD^{(+)}_m \cU_n(\vn)\Big) + \frac{1}{2} \Big(\cDb^{(-)}_{m} \cU_{m}(\vn)\Big)^2\Big]\nn \\
&=& \frac{1}{g^2}\sum_{\vn} \Tr~ \Big[\Big(\cUb_n(\vn + \hatbnu_m)\cUb_m(\vn) - \cUb_m(\vn + \hatbnu_n)\cUb_n(\vn)\Big)\nn \\
&&~~~~~~~~~~~~~~~~\times \Big(\cU_{m}(\vn)\cU_{n}(\vn + \hatbnu_m) - \cU_n(\vn)\cU_m(\vn + \hatbnu_n)\Big) \nn \\
&&~~~~~~~~~~~~~~~~+ \hf \Big(\cU_m(\vn)\cUb_m(\vn) - \cUb_m(\vn - \hatbnu_m)\cU_m(\vn - \hatbnu_m)\Big)^2\Big],
\eea
and the fermionic part
\bea
\label{eq:3d-fermionic-adj}
S_F &=& -\frac{1}{g^2}\sum_{\vn} \Tr~ \Big\{\frac{1}{2}(\delta_{mq}\delta_{nr} - \delta_{mr}\delta_{nq}) \nn \\
&&~~~~~~~~~~~~~~~~ \times \Big[\chi_{mn}(\vn)\Big(\cU_q(\vn)\psi_r(\vn + \hatbnu_q) - \psi_r(\vn)\cU_q(\vn + \hatbnu_r)\Big)\Big] \nn \\
&&~~~~~~~~~~~~~~~~+ \eta(\vn)\Big(\psi_m(\vn)\cUb_m(\vn) - \cUb_m(\vn - \hatbnu_m)\psi_m(\vn - \hatbnu_m)\Big)\nn \\
&&~~~~~~~~~~~~~~~~+\frac{1}{3}(\delta_{mr}\delta_{ne}\delta_{qf} + \delta_{qr}\delta_{me}\delta_{nf} + \delta_{nr}\delta_{qe}\delta_{mf})\nn \\
&&~~~~~~~~~~~~~~~~\times \theta_{ref}(\vn)\Big(\chi_{re}(\vn+\hatbnu_f)\cUb_f(\vn) - \cUb_f(\vn+\hatbnu_r+\hatbnu_e)\chi_{re}(\vn)\Big)\Big\}.
\eea

It is easy to see that each term in the lattice action forms a gauge-invariant loop on the lattice. The lattice action is annihilated by the scalar supercharge $\cQ$. 

\section{Three-dimensional $\cN=4$ lattice SYM with fundamental matter}
\label{sec:latt-form-3dSYM-matter}

We are interested in constructing a three-dimensional $\cN=4$ supersymmetric lattice gauge theory with matter fields in the fundamental representation of $U(N)$. The starting point is a supersymmetric quiver lattice gauge theory with $\cN=4$ defined on two identical copies of three-dimensional spacetimes, which we label the $N_1$-lattice and the $N_2$-lattice. (See Fig. \ref{fig:N1N2lattice}.) The gauge group of the quiver theory is $U(N_1) \times U(N_2)$. The lattice variables on the $N_1$-lattice transform in the representation $({\bf adj}, {\bf 1})$, while those on the $N_2$-lattice transform in the representation $({\bf 1}, {\bf adj})$. The matter fields of the theory live on links connecting the two lattice spacetimes. They transform in the bi-fundamental representations $(\Box, \Boxb)$ and $(\Boxb, \Box)$. See Appendix. \ref{sec:latt-quiver} for details of the lattice construction of three-dimensional $\cN=4$ quiver gauge theory.

In order to construct a lattice gauge theory with matter fields in the fundamental representation we truncate a part of the quiver lattice gauge theory. We have the lattice fields $\{\cU_m(\vn)$, $\cUb_m(\vn)$, $\eta(\vn)$, $\psi_m(\vn)$, $\chi_{mn}(\vn)$, $\theta_{mnr}(\vn)\}$ living on the three-dimensional $N_1$-lattice spacetime and $\{\widehat{\cU}_m(\vn)$, $\widehat{\cUb}_m(\vn)$, $\widehat{\eta}(\vn)$, $\widehat{\psi}_m(\vn)$, $\widehat{\chi}_{mn}(\vn)$, $\widehat{\theta}_{mnr}(\vn)\}$ living on the three-dimensional $N_2$-lattice spacetime. The matter fields $\{\phi(\vn)$, $\widehat{\phib}(\vn)$, $\widehat{\varphi}(\vn)$, $\varphib(\vn)$, $\etab(\vn)$, $\widehat{\psib}_m(\vn)$, $\kappab_{mn}(\vn)$, $\widehat{\thetab}_{mnr}(\vn)\}$ and  $\{\widehat{\phi}(\vn)$, $\phib(\vn)$, $\varphi(\vn)$, $\widehat{\varphib}(\vn)$, $\widehat{\etab}(\vn)$, $\psib_m(\vn)$, $\widehat{\kappab}_{mn}(\vn)$, $\thetab_{mnr}(\vn)\}$ live on links connecting the two lattice spactimes. We make the $N_2$-lattice disappear by making the fields living on the $N_2$-lattice non-dynamical by hand. We also make the matter fields decorated with hats non-dynamical, so that we get only one matter multiplet of the three-dimensional $\cN=4$ gauge theory. Thus we have
\bea
&&\widehat{\cU}_m(\vn) = \widehat{\cUb}_m(\vn) = 1, \\
&&\widehat{\eta}(\vn) = \widehat{\psi}_m(\vn) = \widehat{\chi}_{mn}(\vn) = \widehat{\theta}_{mnr}(\vn) =0, \\
&&\widehat{\phi}(\vn) = \widehat{\phib}(\vn) = \widehat{\varphi}(\vn) = \widehat{\varphib}(\vn) = 0, \\
&&\widehat{\etab}(\vn) = \widehat{\psib}_m(\vn) = \widehat{\kappab}_{mn}(\vn) = \widehat{\thetab}_{mnr}(\vn) = 0.
\eea  
The resultant theory is still supersymmetric, respecting scalar supersymmetry on the lattice. The scalar supersymmetry acts on the lattice fields the following way
\begin{align}
\label{eq:susy-3d-latt-1}
\cQ \cU_m(\vn)& =\psi_m(\vn),&
\cQ \cUb_m(\vn)& =0,\\
\cQ \eta(\vn)& = \Big(\cDb^{(-)}_m \cU_m\Big)(\vn) + (\varphib\varphi)(\vn) - (\phi\phib)(\vn),&
\cQ \psi_m(\vn)& =0,\\
\cQ \chi_{mn}(\vn) & =-\Big(\overline{\cD^{(+)}_m\cU_n}\Big)(\vn),&
\cQ \theta_{mnr}(\vn)& =\epsilon_{mnr} (\varphib\phib)(\vn),\\
\cQ \phi(\vn)& =\etab(\vn),&
\cQ \phib(\vn)& =0,\\
\cQ \varphi(\vn)& =\epsilon_{mnr}\thetab_{mnr}(\vn),&
\cQ \varphib(\vn)& =0,\\
\cQ \etab(\vn)& =0,&
\cQ \psib_m(\vn)& = \Big(\cDb^{(+)}_m\phib\Big)(\vn),\\
\label{eq:susy-3d-latt-2}
\cQ \kappab_{mn}(\vn) & =\epsilon_{mnr} \Big(\cDb^{(+)}_r\varphib\Big)(\vn),&
\cQ \thetab_{mnr}(\vn)& =0.
\end{align}
The action of the resultant theory contains three pieces
\beq
S = S^{\rm bosonic} + S^{\rm fermionic} + S^{\rm matter},
\eeq
where $S^{\rm bosonic}$ and $S^{\rm fermionic}$ contain adjoint fields and the expressions are given in Eq. (\ref{eq:3d-bosonic-adj}) and Eq. (\ref{eq:3d-fermionic-adj}) respectively. The piece $S^{\rm matter}$ contains matter fields in the fundamental representation. It is given by 

\bea
S^{\rm matter} &=& \frac{1}{g^2} \sum_{\vn} \Tr~ \Big\{\Big(\cDb^{(-)}_m\cU_m(\vn)\Big)\Big(\varphib(\vn) \varphi(\vn) - \phi(\vn) \phib(\vn) \Big) \nn \\
&&+ 2 \varphi(\vn) \cD^{(+)}_m\cDb^{(-)}_m \varphib(\vn) + 2 \phib(\vn) \cDb^{(-)}_m\cD^{(+)}_m \phi(\vn) \nn \\
&&+ 2 \psib_m(\vn) \cD^{(+)}_m \etab(\vn) - 2 \kappab_{np}(\vn)\cDb^{(+)}_p \psib_n(\vn) + 2 \thetab_{npm}(\vn)\cD^{(+)}_m \kappab_{np}(\vn) \nn \\
&&+ \eta(\vn) \etab(\vn)\phib(\vn) - \epsilon_{npm} \eta(\vn) \varphib(\vn) \thetab_{npm}(\vn) \nn \\
&&- \frac{1}{3} \epsilon_{mnr} \theta_{mnr}(\vn) \etab(\vn)\varphi(\vn) - 2\psi_m(\vn) \phi(\vn + \hatbnu_m)\psib_m(\vn) \nn \\
&&+ \frac{1}{3}\epsilon_{npm} \psi_m(\vn + \hatbnu_m) \kappab_{np}(\vn + \hatbnu_m)\varphi(\vn) \nn \\
&&+ \epsilon_{npq} \chi_{pq}(\vn + \hatbnu_p + \hatbnu_q) \varphib(\vn)\psib_n(\vn + \hatbnu_p + \hatbnu_q) \nn \\
&&+ \chi_{pq}(\vn +\hatbnu_p + \hatbnu_q) \kappab_{pq}(\vn)\phib(\vn + \hatbnu_p + \hatbnu_q) \nn \\
&&- 2\theta_{mnr}(\vn) \phi(\vn + \hatbnu_m + \hatbnu_n + \hatbnu_r) \thetab_{mnr}(\vn) \nn \\
&&+ \hf \Big(\varphib(\vn)\varphi(\vn) - \phi(\vn)\phib(\vn)\Big)^2 + 2\Big(\varphib(\vn)\phib(\vn) \Big)\Big(\phi(\vn)\varphi(\vn)\Big) \nn \\
&&+ \hf \Big(\phib(\vn)\phi(\vn) - \varphi(\vn)\varphib(\vn)\Big)^2 + 2\Big(\phib(\vn)\varphib(\vn)\Big)\Big(\varphi(\vn)\phi(\vn) \Big)\Big\}.~~~~~~~~~~
\eea

The covariant forward difference operator acts on the lattice variables in the fundamental representation the following way:
\bea
\cD_b^{(+)}f^{(+)\Box}_{a_1 \cdots a_p}(\vn) &\equiv& \cU_b(\vn)f^{(+)\Box}_{a_1 \cdots a_p}(\vn + \hatbnu_b), \\
\cD_b^{(+)}f^{(-)\Box}_{a_1 \cdots a_p}(\vn) &\equiv& \cU_b(\vn+\hatbnu)f^{(-)\Box}_{a_1 \cdots a_p}(\vn + \hatbnu_b), \\
\cDb_b^{(+)}f^{(+)\Box}_{a_1 \cdots a_p}(\vn) &\equiv& -\cUb_b(\vn)f^{(+)\Box}_{a_1 \cdots a_p}(\vn), \\
\cDb_b^{(+)}f^{(-)\Box}_{a_1 \cdots a_p}(\vn) &\equiv& -\cUb_b(\vn+\hatbnu)f^{(-)\Box}_{a_1 \cdots a_p}(\vn).
\eea
The action of the covariant backward difference operator is 
\bea
\cD_b^{(-)}f^{(\pm)\Box}_{a_1 \cdots a_p}(\vn) &\equiv& \cD_b^{(+)}f^{(\pm)\Box}_{a_1 \cdots a_p}(\vn-\hatbnu_b), \\
\cDb_b^{(-)}f^{(\pm)\Box}_{a_1 \cdots a_p}(\vn) &\equiv& \cDb_b^{(+)}f^{(\pm)\Box}_{a_1 \cdots a_p}(\vn-\hatbnu_b).
\eea

For lattice variables in the anti-fundamental representation we have the following set of rules for the action of the covariant forward difference operator
\bea
\cD_b^{(+)}f^{(+)\Boxb}_{a_1 \cdots a_p}(\vn) &\equiv& -f^{(+)\Boxb}_{a_1 \cdots a_p}(\vn) \cU_b(\vn+\hatbnu), \\
\cD_b^{(+)}f^{(-)\Boxb}_{a_1 \cdots a_p}(\vn) &\equiv& -f^{(-)\Boxb}_{a_1 \cdots a_p}(\vn) \cU_b(\vn),\\
\cDb_b^{(+)}f^{(+)\Boxb}_{a_1 \cdots a_p}(\vn) &\equiv& f^{(+)\Boxb}_{a_1 \cdots a_p}(\vn + \hatbnu_b)\cUb_b(\vn+\hatbnu), \\
\cDb_b^{(+)}f^{(-)\Boxb}_{a_1 \cdots a_p}(\vn) &\equiv& f^{(-)\Boxb}_{a_1 \cdots a_p}(\vn + \hatbnu_b)\cUb_b(\vn).
\eea
The covariant backward difference operator acts on the fields the following way
\bea
\cD_b^{(-)}f^{(\pm)\Boxb}_{a_1 \cdots a_p}(\vn) &\equiv& \cD_b^{(+)}f^{(\pm)\Boxb}_{a_1 \cdots a_p}(\vn-\hatbnu_b), \\
\cDb_b^{(-)}f^{(\pm)\Boxb}_{a_1 \cdots a_p}(\vn) &\equiv& \cDb_b^{(+)}f^{(\pm)\Boxb}_{a_1 \cdots a_p}(\vn-\hatbnu_b).
\eea

The fields in the fundamental or anti-fundamental representations are mapped on to lattice sites, with the gauge transformations
\bea
f^{\Box}_{a_1 \cdots a_p}(\vn) &\rightarrow& G(\vn)f^{\Box}_{a_1 \cdots a_p}(\vn), \\
f^{\Boxb}_{a_1 \cdots a_p}(\vn) &\rightarrow& f^{\Boxb}_{a_1 \cdots a_p}(\vn)G^\dagger(\vn).
\eea

Though we have constructed the lattice action for three-dimensional $\cN=4$ lattice gauge theory with fundamental matter using the method of twisting and geometric discretization, we expect that identical lattice can be obtained by the method of orbifold projection. There, the starting point would be the matrix model with sixteen supercharges. The equivalence of the two constructions, for the case of the two-dimensional $\cN=(2, 2)$ SYM lattice gauge theory with fundamental matter, has been proved in Ref. \cite{Matsuura:2008cfa}. 

\section{Renormalization and simulation on the lattice}
\label{sec:renorm-simulation}

In this paper, we have written down lattice actions of three-dimensional $\cN=4$ gauge theories with matter fields. On the lattice, radiative corrections could induce dangerous operators that could violate Lorentz and supersymmetry invariance of the theory as we take the continuum limit. We would like to know whether the above constructed supersymmetric lattice theories are free from fine tuning as the continuum limit is approached. Since three-dimensional gauge theories in general are super-renormalizable, the number of operators that need to be fine-tuned is finite but might not be zero. We could check it at least perturbatively by using a power counting analysis. For a more detailed analysis one has to derive the propagators and vertices of the lattice theory and use lattice perturbation theory to study the radiative corrections.

The counterterms permitted on the lattice are very restrictive - they have to respect the $\cQ$-supersymmetry, gauge symmetry and point group symmetry, $S_3$, on the lattice. The counterterms in the lattice action can take the following generic form, for a given operator $O^{(p)}$ with mass dimension $p$:
\beq
\delta S = \frac{1}{g^2} \int d^3x~ C_p O^{(p)},
\eeq
where $g$ is the coupling parameter, which has mass dimension 1/2. The coefficient $C_p$ denotes the contributions from the loop expansion
\beq
C_p = a^{p - 4} \sum_l c_l (g^2 a)^l,
\eeq
with $l$ counting the number of loops in a perturbative expansion and $a$ denoting the lattice spacing. The dimensionless coefficient $c_l$ can depend at most logarithmically on the lattice spacing - it could be of the form $\log({\bf p} a)$ or $\log(\mu a)$, with ${\bf p}$ and $\mu$ being the external momentum and a mass regulator respectively. Assigning the following mass dimensions to the fields $[\Phi] = 1$, $[\Psi] = \frac{3}{2}$ and $[\cQ] = \hf$, where $\Phi$ and $\Psi$ denote the twisted bosons and fermions respectively, we have the following generic forms for the operators
\bea
O_{\cQ-{\rm exact}}^{(p)} &=& \cQ \Tr \Big(f(\Phi) g(\Psi)\Big), \\
O_{\cQ-{\rm closed}}^{(p)} &=& \Tr \Big(u(\Phi) v(\Psi)\Big).
\eea
We see that such operators are annihilated by the $\cQ$ supersymmetry and thus they are $\cQ$-invariant. Gauge-invariance on the lattice requires that all the fields in the operator must be oriented such that the operator should correspond to the trace of a closed loop on the lattice.

The coefficient of any dangerous operator should vanish in the limit $a \rightarrow 0$, otherwise the continuum limit of the theory would be disastrous. Since radiative corrections start at one-loop ($l=1$) we need to check whether the lattice theory allows operators that respect all the lattice symmetries and with mass dimension $p \leq 3$. Naively we can write down the following set of such operators
\beq
\{ \Phi, \Phi^2, \Phi^3, (\Phi \partial \Phi), (\Psi\Psi), \cQ (\Phi \Psi) \}.
\eeq

Among the set of possible operators, interestingly, we see that the lattice theory has scalar mass terms, $\Tr~\Phi^2$, induced via radiative corrections. For a dimension 2 operator, we have
\beq
\delta S^{(p=2)} = \int d^3x \Big( \frac{c_1}{a} + c_2 g^2 + c_3 g^4 a + \cdots \Big) O^{(p=2)}.
\eeq
The scalar mass terms are thus induced at one- and two-loop, which can have at most a logarithmic divergence, and higher loop vanishing contributions. 

We see that dimension 3 operators, including the fermionic mass terms $\Tr~(\Psi\Psi)$, could be induced at one-loop
\beq
\delta S^{(p=3)} = \int d^3x \Big( c_1 + c_2 g^2a + \cdots \Big) O^{(p=3)},
\eeq
and with vanishing contributions at higher loops. It appears that fermion bilinear counterterms cannot respect all the symmetries of the lattice theory and we conclude that they cannot be generated radiatively.

The lattice theories we constructed above exhibit flat directions (a general feature of theories with extended supersymmetry) and they give rise to instabilities while performing lattice simulations. A way to control them in the simulations is to introduce suitable mass terms to the scalar fields {\it by hand} and then appropriately tune the mass parameters. Since the three-dimensional theories we discussed above can also have scalar mass terms as counterterms we have to tune the bare mass parameters so that they cancel the quantum corrections. These theories might also suffer from a potential sign problem with the fermion determinant. One has to explore the existence of sign problem in these theories to boost confidence that these lattice formulations can be used successfully to explore non-perturbative aspects of three-dimensional $\cN=4$ gauge theories with and without matter\footnote{In Ref. \cite{Catterall:2011aa} it has been shown through lattice simulations that the four and sixteen supercharge SYM theories in two dimensions do not suffer from fermion sign problem.}.

\section{Conclusions and discussion}
\label{sec:conclusion}

In this paper our main result is the lattice formulation of three-dimensional $\cN=4$ supersymmetric gauge theory coupled with matter fields in the fundamental representation. We also briefly discussed the construction of three-dimensional $\cN=4$ gauge theory with adjoint matter. The theory with fundamental matter is formulated in the following way: After dimensionally reducing the twisted version of the four-dimensional $\cN=4$ SYM down to three dimensions, the resultant theory is elevated to $\cN=4$ quiver lattice gauge theory with gauge group $U(N_1) \times U(N_2)$. The method of geometric discretization is used to construct the lattice theory. This lattice theory contains two three-dimensional spacetime lattices, as schematically given in Fig. \ref{fig:N1N2lattice}, which we identify as the $N_1$-lattice and the $N_2$-lattice, with adjoint fields living on the p-cells of each lattice spacetime. The bi-fundamental fields of the theory live on links connecting the two spacetime lattices. The quiver lattice gauge theory is then truncated to a lattice gauge theory on three-dimensional spacetime lattice with gauge group $U(N_1)$. The link fields connecting the two spacetime lattices become matter fields in the fundamental representation and they live on the sites of the $N_1$-lattice. The lattice theory constructed this way is gauge-invariant, free from fermion doubling problem and respects one supersymmetry exactly on the lattice.   

We expect that the same lattice spacetime structure should arise from the method of orbifold projection, of a matrix theory with sixteen supercharges, if one follows the orbifold construction details given in Ref. \cite{Matsuura:2008cfa}, extended to the case of three-dimensional $\cN=4$ gauge theory. 

The lattice theories constructed here admit flat directions, a common feature of theories with extended supersymmetry, and one has to introduce scalar mass terms with tunable mass parameters to simulate them on the lattice. There is a finite set of operators that appear as counterterms in the lattice action including the scalar mass terms, which have to be tuned at two-loop. Such operators have to be carefully studied and enumerated to address the fine tuning issues of these theories on the lattice before embarking on numerical simulations. Study of upto two-loop lattice perturbation theory of these lattice theories is also needed to gain full control over the identification and enumeration of possible counterterms. 

Another pressing question is whether or not these theories suffer from the sign problem of the fermion determinant. One has to perform simulations to measure the phase of the Pfaffian occurring in these theories. In the lattice theories discussed above only one supersymmetry is preserved exactly on the lattice -- the remaining seven supersymmetries are broken by terms of $O(a)$. One has to check whether these remaining supersymmetries are regained in the continuum limit $a \rightarrow 0$ and, if not, how much tuning of the couplings in the lattice action is required\footnote{In Ref. \cite{Catterall:2013roa} it has been argued that restoration of rotational symmetry in the continuum limit of the four-dimensional $\cN=4$ lattice SYM likely implies restoration of R-symmetry and hence should lead to an automatic enhancement to the full $\cN=4$ supersymmetry without further fine-tuning.}.

It would be interesting to look at the three-dimensional $\cN=4$ quiver lattice gauge theory given in Appendix \ref{sec:latt-quiver} in the context of intersecting branes. Three-dimensional $\cN=4$ quiver gauge theories admit realization as low-energy limit brane configurations of Hanany-Witten type \cite{Hanany:1996ie} in type IIB string theory. The string theory contains D3 branes that are stretched between NS5 and D5 branes such that the fivebranes have two common world-volume directions and one common transverse direction. The D3 brane is wrapped on the two common world-volume directions and the common transverse direction. In the field theory limit and at energy scales below the scale set by the interval between the NS5 branes, the world-volume theories on the D3 branes become three-dimensional $U(N)$ $\cN=4$ SYM gauge theories, giving rise to the desired quiver. Three-dimensional $\cN=4$ gauge theories also play an important role in our understanding of dualities. First examples of three-dimensional mirror symmetry \cite{Intriligator:1996ex, Hanany:1996ie, deBoer:1996mp, deBoer:1996ck, Kapustin:1999ha} were provided by such theories and we hope that the lattice theory constructed here would be useful for non-perturbative investigations related to such dualities.

\acknowledgments

I gratefully acknowledge fruitful discussions with Simon Catterall, Sumit R. Das, David B. Kaplan and especially with So Matsuura. This work was supported in part by the LDRD program at the Los Alamos National Laboratory.

\appendix

\section{Lattice formulation of three-dimensional $\cN=4$ quiver theory}
\label{sec:latt-quiver}

In this appendix, we provide the construction details of the three-dimensional $\cN=4$ $U(N_1) \times U(N_2)$ quiver lattice gauge theory coupled with matter fields in the bi-fundamental representation. There are two three-dimensional lattice spacetimes with gauge groups $U(N_1)$ and $U(N_2)$, which we label the $N_1$-lattice and $N_2$-lattice respectively. We denote the position on the $N_1$-lattice by an integer valued three vector $\vn$ while the same position on the $N_2$-lattice is denoted by the vector $\uvn$. 

The lattice fields of the three-dimensional $\cN=4$ SYM are distributed as two identical copies on the two lattice spacetimes. They are given in Table. \ref{tab:1}. The fields on the $N_1$-lattice transform as $({\bf adj}, {\bf 1})$ while those on the $N_2$-lattice transform as $({\bf 1}, {\bf adj})$ under the gauge group $U(N_1) \times U(N_2)$.

The action of the forward and backward covariant difference operators on fields living on $N_1$-lattice is summarized in Eqs. (\ref{eq:cov-diff-adj-1}) - (\ref{eq:cov-diff-adj-2}). 

\begin{figure}
\begin{center}
\includegraphics[width=0.6\textwidth]{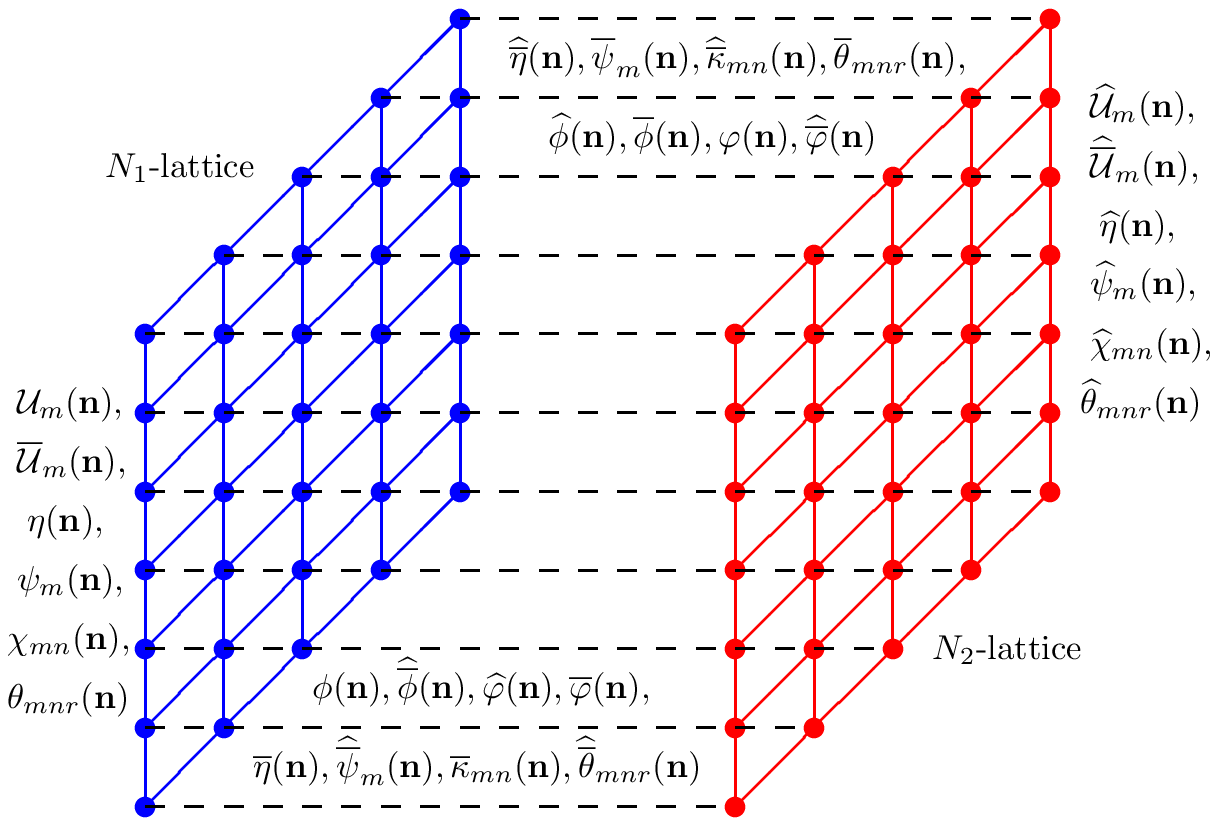}
\end{center}
\caption{\label{fig:N1N2lattice}Schematic sketch of the lattice construction of three-dimensional $\cN=4$ quiver gauge theory. The lattice variables $\{\cU_m(\vn)$, $\cUb_m(\vn)$, $\eta(\vn)$, $\psi_m(\vn)$, $\chi_{mn}(\vn)$, $\theta_{mnr}(\vn)\}$ live on the three-dimensional $N_1$-lattice spacetime and $\{\widehat{\cU}_m(\vn)$, $\widehat{\cUb}_m(\vn)$, $\widehat{\eta}(\vn)$, $\widehat{\psi}_m(\vn)$, $\widehat{\chi}_{mn}(\vn)$, $\widehat{\theta}_{mnr}(\vn)\}$ live on the three-dimensional $N_2$-lattice spacetime. The matter fields $\{\phi(\vn)$, $\widehat{\phib}(\vn)$, $\widehat{\varphi}(\vn)$, $\varphib(\vn)$, $\etab(\vn)$, $\widehat{\psib}_m(\vn)$, $\kappab_{mn}(\vn)$, $\widehat{\thetab}_{mnr}(\vn)\}$ and  $\{\widehat{\phi}(\vn)$, $\phib(\vn)$, $\varphi(\vn)$, $\widehat{\varphib}(\vn)$, $\widehat{\etab}(\vn)$, $\psib_m(\vn)$, $\widehat{\kappab}_{mn}(\vn)$, $\thetab_{mnr}(\vn)\}$ live on links connecting the two lattice spactimes.}
\end{figure}

The matter fields of the quiver lattice theory live on links connecting the $N_1$-lattice and the $N_2$-lattice spacetimes. They are in the bi-fundamental representations of $U(N_1) \times U(N_2)$. In Tab. \ref{tab:2} we provide the set of matter fields and their representations. In Fig. \ref{fig:N1N2lattice} we summarize the field content and structure of the three-dimensional $\cN=4$ quiver lattice gauge theory.

We need to define the action of the covariant difference operators on the lattice fields in the bi-fundamental representations. We have the following set of rules.

For lattice variables in the representation $(\Box, \Boxb)$ the covariant forward difference operator acts the following way:
\bea
&&\cD_b^{(+)}f^{(+)}_{a_1 \cdots a_p}(\vn, \uvn) \equiv \cU_b(\vn)f^{(+)}_{a_1 \cdots a_p}(\vn + \hatbnu_b, \uvn + \uhatbnu_b)-f^{(+)}_{a_1 \cdots a_p}(\vn, \uvn) \widehat{\cU}_b(\uvn),~~~~~~~~ \\
&&\cD_b^{(+)}f^{(-)}_{a_1 \cdots a_p}(\vn, \uvn) \equiv \cU_b(\vn+\hatbnu)f^{(-)}_{a_1 \cdots a_p}(\vn + \hatbnu_b, \uvn + \uhatbnu_b)-f^{(-)}_{a_1 \cdots a_p}(\vn, \uvn) \widehat{\cU}_b(\uvn),~~~~~~~~ \\
&&\cDb_b^{(+)}f^{(+)}_{a_1 \cdots a_p}(\vn, \uvn) \equiv f^{(+)}_{a_1 \cdots a_p}(\vn + \hatbnu_b, \uvn + \uhatbnu_b)\widehat{\cUb}_b(\uvn)-\cUb_b(\vn)f^{(+)}_{a_1 \cdots a_p}(\vn, \uvn),~~~~~~~~ \\
&&\cDb_b^{(+)}f^{(-)}_{a_1 \cdots a_p}(\vn, \uvn) \equiv f^{(-)}_{a_1 \cdots a_p}(\vn + \hatbnu_b, \uvn + \uhatbnu_b)\widehat{\cUb}_b(\uvn)-\cUb_b(\vn+\hatbnu)f^{(-)}_{a_1 \cdots a_p}(\vn, \uvn),
\eea
while the covariant backward difference operator acts on the fields according to the rules
\bea
\cD_b^{(-)}f^{(\pm)}_{a_1 \cdots a_p}(\vn, \uvn) &\equiv& \cD_b^{(+)}f^{(\pm)}_{a_1 \cdots a_p}(\vn-\hatbnu_b, \uvn-\uhatbnu_b), \\
\cDb_b^{(-)}f^{(\pm)}_{a_1 \cdots a_p}(\vn, \uvn) &\equiv& \cDb_b^{(+)}f^{(\pm)}_{a_1 \cdots a_p}(\vn-\hatbnu_b, \uvn-\uhatbnu_b).
\eea

For lattice variables in the representation $(\Boxb, \Box)$ we have the following set of rules for the covariant difference operators:
\bea
&&\cD_b^{(+)}f^{(+)}_{a_1 \cdots a_p}(\uvn, \vn) \equiv \widehat{\cU}_b(\uvn)f^{(+)}_{a_1 \cdots a_p}(\uvn + \uhatbnu_b, \vn + \hatbnu_b)-f^{(+)}_{a_1 \cdots a_p}(\uvn, \vn) \cU_b(\vn),~~~~~~~~ \\
&&\cD_b^{(+)}f^{(-)}_{a_1 \cdots a_p}(\uvn, \vn) \equiv \widehat{\cU}_b(\uvn+\uhatbnu)f^{(-)}_{a_1 \cdots a_p}(\uvn + \uhatbnu_b, \vn + \hatbnu_b)-f^{(-)}_{a_1 \cdots a_p}(\uvn, \vn) \cU_b(\vn),~~~~~~~~ \\
&&\cDb_b^{(+)}f^{(+)}_{a_1 \cdots a_p}(\uvn, \vn) \equiv f^{(+)}_{a_1 \cdots a_p}(\uvn + \uhatbnu_b, \vn + \hatbnu_b)\cUb_b(\vn)-\widehat{\cUb}_b(\uvn)f^{(+)}_{a_1 \cdots a_p}(\uvn, \vn),~~~~~~~~ \\
&&\cDb_b^{(+)}f^{(-)}_{a_1 \cdots a_p}(\uvn, \vn) \equiv f^{(-)}_{a_1 \cdots a_p}(\uvn + \uhatbnu_b, \vn + \hatbnu_b)\cUb_b(\vn)-\widehat{\cUb}_b(\uvn+\uhatbnu)f^{(-)}_{a_1 \cdots a_p}(\uvn, \vn),
\eea
and
\bea
\cD_b^{(-)}f^{(\pm)}_{a_1 \cdots a_p}(\uvn, \vn) &\equiv& \cD_b^{(+)}f^{(\pm)}_{a_1 \cdots a_p}(\uvn-\uhatbnu_b, \vn-\hatbnu_b), \\
\cDb_b^{(-)}f^{(\pm)}_{a_1 \cdots a_p}(\uvn, \vn) &\equiv& \cDb_b^{(+)}f^{(\pm)}_{a_1 \cdots a_p}(\uvn-\uhatbnu_b, \vn-\hatbnu_b).
\eea

The action of the three-dimensional $\cN=4$ quiver lattice theory contains the following pieces
\beq
S = S^{\rm bosonic}_{({\bf adj}, {\bf 1})} + S^{\rm bosonic}_{({\bf 1}, {\bf adj})} + S^{\rm fermionic}_{({\bf adj}, {\bf 1})} + S^{\rm fermionic}_{({\bf 1}, {\bf adj})} + S^{\rm matter}_{(\Box, \Boxb)} + S^{\rm matter}_{(\Boxb, \Box)},
\eeq
where
\bea
S^{\rm bosonic}_{({\bf adj}, {\bf 1})} &=& \frac{1}{g^2}\sum_{\vn} \Tr~ \Big[\Big(\cUb_n(\vn + \hatbnu_m)\cUb_m(\vn) - \cUb_m(\vn + \hatbnu_n)\cUb_n(\vn)\Big)\nn \\
&&~~~~~~~~~~~~~~~~~~\times \Big(\cU_{m}(\vn)\cU_{n}(\vn + \hatbnu_m) - \cU_n(\vn)\cU_m(\vn + \hatbnu_n)\Big)\nn \\
&&~~~~~~~~~~~~~~~~~~+ \hf \Big(\cU_m(\vn)\cUb_m(\vn) - \cUb_m(\vn - \hatbnu_m)\cU_m(\vn - \hatbnu_m)\Big)^2\Big],
\eea
\bea
S^{\rm bosonic}_{({\bf 1}, {\bf adj})} &=& \frac{1}{g^2}\sum_{\uvn} \Tr~ \Big[\Big(\widehat{\cUb}_n(\uvn + \uhatbnu_m)\widehat{\cUb}_m(\uvn) - \widehat{\cUb}_m(\uvn + \uhatbnu_n)\widehat{\cUb}_n(\uvn)\Big)\nn \\
&&~~~~~~~~~~~~~~~~~~\times \Big(\widehat{\cU}_{m}(\uvn)\widehat{\cU}_{n}(\uvn + \uhatbnu_m) - \widehat{\cU}_n(\uvn)\widehat{\cU}_m(\uvn + \uhatbnu_n)\Big)\nn \\
&&~~~~~~~~~~~~~~~~~~+ \hf \Big(\widehat{\cU}_m(\uvn)\widehat{\cUb}_m(\uvn) - \widehat{\cUb}_m(\uvn - \uhatbnu_m)\widehat{\cU}_m(\uvn - \uhatbnu_m)\Big)^2\Big],
\eea
\bea
S^{\rm fermionic}_{({\bf adj}, {\bf 1})} &=& -\frac{1}{g^2}\sum_{\vn} \Tr~ \Big\{\hf(\delta_{mq}\delta_{nr} - \delta_{mr}\delta_{nq}) \nn \\
&&~~~~~~~~\times \Big[\chi_{mn}(\vn)\Big(\cU_q(\vn)\psi_r(\vn + \hatbnu_q) - \psi_r(\vn)\cU_q(\vn + \hatbnu_r)\Big)\Big] \nn \\
&&~~~~~~~~+ \eta(\vn)\Big(\psi_m(\vn)\cUb_m(\vn) - \cUb_m(\vn - \hatbnu_m)\psi_m(\vn - \hatbnu_m)\Big)\nn \\
&&~~~~~~~~+\frac{1}{3}(\delta_{mr}\delta_{ne}\delta_{qf} + \delta_{qr}\delta_{me}\delta_{nf} + \delta_{nr}\delta_{qe}\delta_{mf})\nn \\
&&~~~~~~~~\times \theta_{ref}(\vn)\Big(\chi_{re}(\vn+\hatbnu_f)\cUb_f(\vn) - \cUb_f(\vn+\hatbnu_r+\hatbnu_e)\chi_{re}(\vn)\Big)\Big\},
\eea

\begin{table}[tbp]
\centering
\begin{tabular}{| c | c | c |}
\hline
  Field & $N_1$-lattice & $N_2$-lattice \\
\hline \hline
$\cU_m$ & $\cU_m(\vn, \vn+\hatbnu_m)$ & $\widehat{\cU}_m(\uvn, \uvn+\uhatbnu_m)$ \\
\hline
$\cUb_m$ & $\cUb_m(\vn+\hatbnu_m, \vn)$ & $\widehat{\cUb}_m(\uvn+\uhatbnu_m, \uvn)$ \\
\hline
$\eta$ & $\eta(\vn, \vn)$ & $\widehat{\eta}(\uvn, \uvn)$ \\
\hline
$\psi_m$ & $\psi_m(\vn, \vn+\hatbnu_m)$ & $\widehat{\psi}_m(\uvn, \uvn+\uhatbnu_m)$ \\
\hline
$\chi_{mn}$ & $\chi_{mn}(\uvn+\uhatbnu_m+\uhatbnu_n, \vn)$ & $\widehat{\chi}_{mn}(\uvn+\uhatbnu_m+\uhatbnu_n, \uvn)$ \\
\hline
$\theta_{mnr}$ & $\theta_{mnr}(\vn, \vn+\hatbnu_m+\hatbnu_n+\hatbnu_r)$ & $\widehat{\theta}_{mnr}(\uvn, \uvn+\uhatbnu_m+\uhatbnu_n+\uhatbnu_r)$ \\
\hline
\end{tabular}
\caption{\label{tab:1} The placement of adjoint fields of the three-dimensional $\cN=4$ quiver lattice gauge theory.}
\end{table}

\begin{table}[tbp]
\centering
\begin{tabular}{| c | c | c |}
\hline
  Field & $(\Box, \Boxb)$ & $(\Boxb, \Box)$ \\
\hline \hline
$\phi$ & $\phi(\vn, \uvn)$ & $\widehat{\phi}(\uvn, \vn)$ \\
\hline
$\phib$ & $\widehat{\phib}(\vn, \uvn)$ & $\phib(\uvn, \vn)$ \\
\hline
$\varphi$ & $\widehat{\varphi}(\vn, \uvn)$ & $\varphi(\uvn, \vn)$ \\
\hline
$\varphib$ & $\varphib(\vn, \uvn)$ & $\widehat{\varphib}(\uvn, \vn)$ \\
\hline
$\etab$ & $\etab(\vn, \uvn)$ & $\widehat{\etab}(\uvn, \vn)$ \\
\hline
$\psib_m$ & $\widehat{\psib}_m(\vn+\hatbnu_m, \uvn)$ & $\psib_m(\uvn+\uhatbnu_m, \vn)$ \\
\hline
$\kappab_{mn}$ & $\kappab_{mn}(\vn+\hatbnu_m+\hatbnu_n, \uvn)$ & $\widehat{\kappab}_{mn}(\uvn+\uhatbnu_m+\uhatbnu_n, \vn)$ \\
\hline
$\thetab_{mnr}$ & $\widehat{\thetab}_{mnr}(\vn+\hatbnu_m+\hatbnu_n+\hatbnu_r, \uvn)$ & $\thetab_{mnr}(\uvn+\uhatbnu_m+\uhatbnu_n+\uhatbnu_r, \vn)$ \\
\hline
\end{tabular}
\caption{\label{tab:2} The placement of matter fields of the three-dimensional $\cN=4$ quiver lattice gauge theory. They transform in the bi-fundamental representations of $U(N_1) \times U(N_2)$.}
\end{table}

\bea
S^{\rm fermionic}_{({\bf 1}, {\bf adj})} &=& -\frac{1}{g^2}\sum_{\uvn} \Tr~ \Big\{\hf(\delta_{mq}\delta_{nr} - \delta_{mr}\delta_{nq}) \nn \\
&&~~~~~~~~\times \Big[\widehat{\chi}_{mn}(\uvn)\Big(\widehat{\cU}_q(\uvn)\widehat{\psi}_r(\uvn + \uhatbnu_q) - \widehat{\psi}_r(\uvn)\widehat{\cU}_q(\uvn + \uhatbnu_r)\Big)\Big] \nn \\
&&~~~~~~~~+ \widehat{\eta}(\uvn)\Big(\widehat{\psi}_m(\uvn)\widehat{\cUb}_m(\uvn) - \widehat{\cUb}_m(\uvn - \uhatbnu_m)\widehat{\psi}_m(\uvn - \uhatbnu_m)\Big)\nn \\
&&~~~~~~~~+\frac{1}{3}(\delta_{mr}\delta_{ne}\delta_{qf} + \delta_{qr}\delta_{me}\delta_{nf} + \delta_{nr}\delta_{qe}\delta_{mf})\nn \\
&&~~~~~~~~\times \widehat{\theta}_{ref}(\uvn)\Big(\widehat{\chi}_{re}(\uvn+\uhatbnu_f)\widehat{\cUb}_f(\uvn) - \widehat{\cUb}_f(\uvn+\uhatbnu_r+\uhatbnu_e)\widehat{\chi}_{re}(\uvn)\Big)\Big\},
\eea

\bea
S^{\rm matter}_{(\Box, \Boxb)} &=& \frac{1}{g^2}\sum_{\vn} \Tr~ \Big\{\Big(\cDb^{(-)}_m\cU_m(\vn)\Big)\Big(\varphib(\vn, \uvn) \varphi(\uvn, \vn) - \widehat{\varphi}(\vn, \uvn) \widehat{\varphib}(\uvn, \vn) \nn \\
&&+ \widehat{\phib}(\vn, \uvn) \widehat{\phi}(\uvn, \vn) - \phi(\vn, \uvn) \phib(\uvn, \vn)\Big) \nn \\
&&+ 2 \widehat{\varphi}(\vn, \uvn) \cD^{(+)}_m\cDb^{(-)}_m \widehat{\varphib}(\uvn, \vn) + 2 \widehat{\phib}(\vn, \uvn) \cDb^{(-)}_m\cD^{(+)}_m \widehat{\phi}(\uvn, \vn) \nn \\
&&+ 2\widehat{\psib}_m(\vn + \hatbnu_m, \uvn) \cD^{(+)}_m \widehat{\etab}(\uvn, \vn) + 2 \widehat{\thetab}_{npm}(\vn + \hatbnu_n, \uvn)\cD^{(+)}_m\widehat{\kappab}_{np}(\uvn, \vn) \nn \\
&&- 2 \kappab_{np}(\vn, \uvn + \uhatbnu_p)\cDb^{(+)}_p\psib_n(\uvn, \vn) - \eta(\vn, \vn) \Big(\widehat{\phib}(\vn, \uvn)\widehat{\etab}(\uvn, \vn) - \etab(\vn, \uvn)\phib(\uvn, \vn)\Big) \nn \\
&&- \epsilon_{npm} \eta(\vn, \vn) \Big(\varphib(\vn, \uvn) \thetab_{npm}(\uvn, \vn) - \widehat{\thetab}_{npm}(\vn, \uvn)\widehat{\varphib}(\uvn, \vn)\Big) \nn \\
&&+ \frac{1}{3} \epsilon_{mnr} \theta_{mnr}(\vn, \vn) \Big(\widehat{\varphi}(\vn, \uvn)\widehat{\etab}(\uvn, \vn) - \etab(\vn, \uvn)\varphi(\uvn, \vn)\Big) \nn \\
&&- 2\psi_m(\vn, \vn + \hatbnu_m) \Big(\phi(\vn + \hatbnu_m, \uvn + \uhatbnu_m)\psib_m(\uvn + \uhatbnu_m, \vn) - \widehat{\psib}_m(\vn + \hatbnu_m, \uvn)\widehat{\phi}(\uvn, \vn)\Big) \nn \\
&&- \frac{1}{3}\epsilon_{npm} \psi_m(\vn + \hatbnu_m, \vn) \Big(\widehat{\varphi}(\vn + \hatbnu_m, \uvn + \uhatbnu_m)\widehat{\kappab}_{np}(\uvn + \uhatbnu_m, \vn) \nn \\
&&- \kappab_{np}(\vn + \hatbnu_m, \uvn)\varphi(\uvn, \vn)\Big) \nn \\
&&+ \epsilon_{npq} \chi_{pq}(\vn + \hatbnu_p + \hatbnu_q, \vn) \Big(\varphib(\vn, \uvn + \uhatbnu_p + \uhatbnu_q)\psib_n(\uvn + \uhatbnu_p + \uhatbnu_q, \vn + \hatbnu_p + \hatbnu_q) \nn \\
&&- \widehat{\psib}_n(\vn, \uvn + \uhatbnu_p + \uhatbnu_q)\widehat{\varphib}(\uvn + \uhatbnu_p + \uhatbnu_q, \vn + \hatbnu_p + \hatbnu_q)\Big) \nn \\
&&- \chi_{pq}(\vn +\hatbnu_p + \hatbnu_q, \vn) \Big(\widehat{\phib}(\vn, \uvn)\widehat{\kappab}_{pq}(\uvn, \vn + \hatbnu_p + \hatbnu_q) \nn \\
&&- \kappab_{pq}(\vn, \uvn + \uhatbnu_p + \uhatbnu_q)\phib(\uvn + \uhatbnu_p + \uhatbnu_q, \vn + \hatbnu_p + \hatbnu_q)\Big) \nn \\
&&- 2\theta_{mnr}(\vn, \vn + \hatbnu_m + \hatbnu_n + \hatbnu_r)\Big(\phi(\vn + \hatbnu_m + \hatbnu_n + \hatbnu_r, \uvn + \uhatbnu_m + \uhatbnu_n + \uhatbnu_r)\thetab_{mnr}(\uvn, \vn) \nn \\
&&- \widehat{\thetab}_{mnr}(\vn, \uvn + \uhatbnu_m + \uhatbnu_n + \uhatbnu_r)\widehat{\phi}(\uvn, \vn)\Big) \nn \\
&&+ \hf \Big(\varphib(\vn, \uvn)\varphi(\uvn, \vn) - \widehat{\varphi}(\vn, \uvn)\widehat{\varphib}(\uvn, \vn)+ \widehat{\phib}(\vn, \uvn)\widehat{\phi}(\uvn, \vn) - \phi(\vn, \uvn)\phib(\uvn, \vn)\Big)^2 \nn \\
&&- 2\Big(\varphib(\vn, \uvn)\phib(\uvn, \vn) - \widehat{\phib}(\vn, \uvn)\widehat{\varphib}(\uvn, \vn)\Big)\Big(\widehat{\varphi}(\vn, \uvn)\widehat{\phi}(\uvn, \vn) - \phi(\vn, \uvn)\varphi(\uvn, \vn)\Big)\Big\},~~~~~~~~~~
\eea

\bea
S^{\rm matter}_{(\Boxb, \Box)} &=& \frac{1}{g^2}\sum_{\vn} \Tr~ \Big\{\Big(\cDb^{(-)}_m\widehat{\cU}_m(\uvn)\Big)\Big(\widehat{\varphib}(\uvn, \vn) \widehat{\varphi}(\vn, \uvn) - \varphi(\uvn, \vn) \varphib(\vn, \uvn) \nn \\
&&+ \phib(\uvn, \vn) \phi(\vn, \uvn) - \widehat{\phi}(\uvn, \vn) \widehat{\phib}(\vn, \uvn)\Big) \nn \\
&&+ 2 \varphi(\uvn, \vn) \cD^{(+)}_m\cDb^{(-)}_m \varphib(\vn, \uvn) + 2 \phib(\uvn, \vn) \cDb^{(-)}_m\cD^{(+)}_m \phi(\vn, \uvn) \nn \\
&&+ 2 \psib_m(\uvn + \uhatbnu_m, \vn) \cD^{(+)}_m \etab(\vn, \uvn) + 2 \thetab_{npm}(\uvn + \uhatbnu_n, \vn)\cD^{(+)}_m \kappab_{np}(\vn, \uvn) \nn \\
&&- 2 \widehat{\kappab}_{np}(\uvn, \vn + \hatbnu_p)\cDb^{(+)}_p \widehat{\psib}_n(\vn, \uvn) - \widehat{\eta}(\uvn, \uvn) \Big(\phib(\uvn, \vn) \etab(\vn, \uvn) - \widehat{\etab}(\uvn, \vn)\widehat{\phib}(\vn, \uvn)\Big) \nn \\
&&- \epsilon_{npm} \widehat{\eta}(\uvn, \uvn) \Big(\widehat{\varphib}(\uvn, \vn) \widehat{\thetab}_{npm}(\vn, \uvn) - \thetab_{npm}(\uvn, \vn) \varphib(\vn, \uvn)\Big) \nn \\
&&+ \frac{1}{3} \epsilon_{mnr} \widehat{\theta}_{mnr}(\uvn, \uvn) \Big(\varphi(\uvn, \vn) \etab(\vn, \uvn) - \widehat{\etab}(\uvn, \vn)\widehat{\varphi}(\vn, \uvn)\Big) \nn \\
&&- 2\widehat{\psi}_m(\uvn, \uvn + \uhatbnu_m) \Big(\widehat{\phi}(\uvn + \uhatbnu_m, \vn + \hatbnu_m)\widehat{\psib}_m(\vn + \hatbnu_m, \uvn) - \psib_m(\uvn + \uhatbnu_m, \vn)\phi(\vn, \uvn)\Big) \nn \\
&&- \frac{1}{3}\epsilon_{npm} \widehat{\psi}_m(\uvn + \uhatbnu_m, \uvn) \Big(\varphi(\uvn + \uhatbnu_m, \vn + \hatbnu_m) \kappab_{np}(\vn + \hatbnu_m, \uvn) \nn \\
&&- \widehat{\kappab}_{np}(\uvn + \uhatbnu_m, \vn) \widehat{\varphi}(\vn, \uvn)\Big) \nn \\
&&+ \epsilon_{npq} \widehat{\chi}_{pq}(\uvn + \uhatbnu_p + \uhatbnu_q, \uvn) \Big(\widehat{\varphib}(\uvn, \vn + \hatbnu_p + \hatbnu_q)\widehat{\psib}_n(\vn + \hatbnu_p + \hatbnu_q, \uvn + \uhatbnu_p + \uhatbnu_q) \nn \\
&&- \psib_n(\uvn, \vn + \hatbnu_p + \hatbnu_q)\varphib(\vn + \hatbnu_p + \hatbnu_q, \uvn + \uhatbnu_p + \uhatbnu_q)\Big) \nn \\
&&- \widehat{\chi}_{pq}(\uvn +\uhatbnu_p + \uhatbnu_q, \uvn) \Big(\phib(\uvn, \vn)\kappab_{pq}(\vn, \uvn + \uhatbnu_p + \uhatbnu_q) \nn \\
&&- \widehat{\kappab}_{pq}(\uvn, \vn + \hatbnu_p + \hatbnu_q)\widehat{\phib}(\vn + \hatbnu_p + \hatbnu_q, \uvn + \uhatbnu_p + \uhatbnu_q)\Big) \nn \\
&&- 2\widehat{\theta}_{mnr}(\uvn, \uvn + \uhatbnu_m + \uhatbnu_n + \uhatbnu_r)\Big(\widehat{\phi}(\uvn + \uhatbnu_m + \uhatbnu_n + \uhatbnu_r, \vn + \hatbnu_m + \hatbnu_n + \hatbnu_r)\widehat{\thetab}_{mnr}(\vn, \uvn) \nn \\
&&- \thetab_{mnr}(\uvn, \vn + \hatbnu_m + \hatbnu_n + \hatbnu_r)\phi(\vn, \uvn)\Big) \nn \\
&&+ \hf \Big(\widehat{\varphib}(\uvn, \vn)\widehat{\varphi}(\vn, \uvn) - \varphi(\uvn, \vn)\varphib(\vn, \uvn) + \phib(\uvn, \vn)\phi(\vn, \uvn) - \widehat{\phi}(\uvn, \vn)\widehat{\phib}(\vn, \uvn)\Big)^2 \nn \\
&&- 2\Big(\widehat{\varphib}(\uvn, \vn)\widehat{\phib}(\vn, \uvn) - \phib(\uvn, \vn)\varphib(\vn, \uvn)\Big)\Big(\varphi(\uvn, \vn)\phi(\vn, \uvn) - \widehat{\phi}(\uvn, \vn)\widehat{\varphi}(\vn, \uvn)\Big)\Big\}.~~~~~~~~~~
\eea

\end{document}